# VUV Pump and probe of phase separation and oxygen interstitials in $La_2NiO_{4+y}$ using spectromicroscopy

Antonio Bianconi[1,2,3,4], Augusto Marcelli[1,5], Markus Bendele[1], Davide Innocenti[6], Alexei Barinov[7], Nathalie Poirot[8], Gaetano Campi[2]

[1] *Rome International Center of Materials Science Superstripes, RICMASS, Via dei Sabelli 119A, 00185 Rome, Italy*
[2] *Institute of Crystallography, Consiglio Nazional delle Ricerche, IC-CNR, Via Salaria Km 29.300, Monterotondo, Rome, I-00015, Italy*
[3] *National Research Nuclear University, MEPhI (Moscow Engineering Physics Institute), Kashirskoye sh. 31, Moscow 115409, Russia*
[4] *Latvia Academy of Science, Akadēmijas laukums 1, LV-1050 Riga. Latvia*
[5] *Istituto Nazionale di Fisica Nucleare, Laboratori Nazionali di Frascati, 00044 Frascati, Italy*
[6] *Department of Chemistry, University of Liverpool, UK*
[7] *Sincrotrone Trieste S.C.p.A., Area Science Park, 34012 Basovizza, Trieste, Italy*
[7] *Université de Tours, IUT de Bloism 15 rue de la Chocolaterie, C.S. 2903, 41000 Blois, France*

While it is known that strongly correlated transition metal oxides described by multi-band Hubbard model show microscopic multiscale phase separation little is known on the possibility to manipulate them with vacuum ultraviolet (VUV), 27 eV lighting. We have investigated the photo-induced effects of VUV illumination of a super-oxygenated $La_2NiO_{4+y}$ single crystal by means of scanning photoelectron microscopy. VUV light exposure induces the increase of the density of states (DOS) in the binding energy range around $E_b$ =1.4 eV below $E_F$. The photo-induced states in this energy region have been predicted as due to clustering of oxygen interstitials by band structure calculations for large supercell of $La_2CuO_{4.125}$. We finally show that it possible to generate and manipulate oxygen rich domains by VUV illumination as it was reported for X-ray illumination of $La_2CuO_{4+y}$. This phenomenology is assigned to oxygen-interstitials ordering and clustering by photo-illumination forming segregated domains in the $La_2NiO_{4+y}$ surface.

1. INTRODUCTION

Writing patterns in organic and inorganic media by illumination, starting from silver-halide processes for traditional photography, is a key method to manipulate materials for advanced technologies. In the last decade photo-induced effects have been investigated in the families of strongly correlated complex quantum matter like transition metal oxides, showing high temperature superconductivity [1-14] and colossal magneto resistance [15,16]. Controlling photo-induced effects in complex matter is of high interest in nanotechnology for novel oxide nanoelectronics on demand [17-21]. The interest has been mostly addressed on ($A_2MO_{4+y}$) system having the $K_2NiF_4$-type structure with A=Cu i.e., $La_2CuO_{4+y}$ structure, which received much attention since these compounds show nano-scale phase separation [22-24]. The emergence of multiscale phase separation from nano-scale to micron-scale has been explained to be driven by tuning the chemical potential at a Lifshitz transition in a multi-band Hubbard model [25-27]. In this regime the competition between spin, charge, orbital and elastic interactions can drive the system into metastable phases, i.e., quasi stationary states out of equilibrium with the coexistence and competition between a metallic phase and a localized charge ordered phase. While "large polarons" spanning about 8 lattice sites in the intermediate coupling regime have been found in cuprates, [28] "small polarons" localized in a single lattice in the strong coupling regime have been found in manganites. [29] In these systems a relevant lattice effect is due to mobile oxygen interstitials in the spacer layers, which contribute to the complexity with the formation of dopant rich domains anticorrelated with charge ordered domains which control the nanoelectronic functionality [30].

In these complex systems characterized by a large variety of coexisting superconducting, insulating, ferromagnetic, antiferromagnetic states, X-ray illumination induces electronic and structural changes [1-21] allowing tuning of material functionalities, which allow the development of many device capabilities.

The electronic properties of super-oxygenated $La_2NiO_{4+y}$ have high technological interest [31-39]. This ($A_2MO_{4+y}$) system A=Ni having the $K_2NiF_4$-type structure has the ability to accommodate a large oxygen over-stoichiometry. It is formed by a stack of bcc atomic [$NiO_2$] layers intercalated by [$La_2O_2$] atomic layers similar to the simplest high temperature cuprate superconductor $La_2CuO_{4+y}$, but it does not show superconductivity at any measurable doping level. It may be possible that oxygen interstitial ordering at room temperature can be manipulated in $La_2NiO_{4+y}$ as in the cuprate $La_2CuO_{4+y}$.

The [$NiO_2$] layers are typical charge transfer Mott insulators [31-39]. The $Ni^{2+}$ ion have a Ni $3d^8$ configuration whereas the antiferromagnetic order is comparable to NiO. The mobile oxygen interstitials in $La_2NiO_{4+y}$ enter in the rocksalt spacer layer [$La_2O_{2+y}$] and sit at (1/4,1/4,1/4) type positions of the orthorhombic lattice creating $nh=2y$ holes into the $NiO_2$ planes. The doped holes enter in the oxygen 2p orbital L forming $3d^8L$ localized states, similar to NiO [40]. The $3d^8L$ states form "small polarons" localized on single atomic oxygen L sites in the $NiO_2$ plane where the doped charge is associated with a local lattice distortion of the $NiO_2$ plane similarly to the insulating phase of manganites [29]. The idea of polaron ordering in doped $NiO_2$, as the source of magnetic stripes driven by doping, has been proposed by Zaanen and Littlewood [33]. In doped Mott-Hubbard insulators, the electron-phonon interaction and the strong Coulomb repulsion can reinforce each other to stabilize small polarons, domain walls, and charge-density wave. At low temperature the holes are ordered forming polaronic stripes of localized charges and magnetic moments in the diagonal direction of the Ni-O bond direction. The complexity of the striped magnetic phase is related with phase separation and ordering of oxygen interstitials.

Measurements of the in-plane resistance in the nickelates suggest that oxygen interstitial orderings appear below $T_{CO}$=320 K. In comparison the magnetic and small polaron ordering occurs at temperature lower than $T_m$=110 K.

The ordering of oxygen interstitials in the $CuO_2$ plane in the families of cuprates is controlled by the compressive misfit strain [41,42] in the [$CuO_2$] active layer near the spacer layer [$La_2O_{2+y}$]. The [$CuO_2$] compressive strain is compensated by the [$La_2O_{2+y}$] tensile misfit strain in the formation of the multilayer crystal. The increase of the [$CuO_2$] compressive misfit strain pushes the system to the formation of small polarons, but at the same time the [$La_2O_{2+y}$] tensile misfit strain determines the increase of the mobility of the $y$ oxygen interstitials. This is the case of doped $La_2NiO_{4+y}$ where, due to the large tensile misfit strain in the $[La_2O_{2+y}]$, the mobility of oxygen interstitials is high. The latter is in fact an oxygen ion conductor at high temperature with high technological relevance [43-50]. Synchrotron radiation investigations of $La_2NiO_{4+y}$ [51,52] and $La_2CuO_{4+y}$ [53] have been performed using standard XANES, x-ray absorption near edge structure, [54-58] methods. We underline here that in these system it is necessary to take into consideration many body final states configurations, relevant in all strongly correlated oxides. [59-65]

2.  RESULTS AND DISCUSSION

Scanning photoelectron microscopy (SPEM) was performed on the insulating $La_2NiO_{4+y}$ (y=0.14) [34,35,66]. The SPEM measurements were performed at the spectromicroscopy beamline at the Elettra synchrotron light source, Trieste (Italy) [67,68]. Data presented here were collected at the photon energy of 27 eV. The beam size in focus is ~500 nm FWHM by a Schwarzschild objective. The VUV illumination was performed by leaving sample under same but defocused beam having donut like shape due to a presence of central stop in the optics. Surface effects such as topographical contrast were eliminated by following the procedure described in Ref. [67].

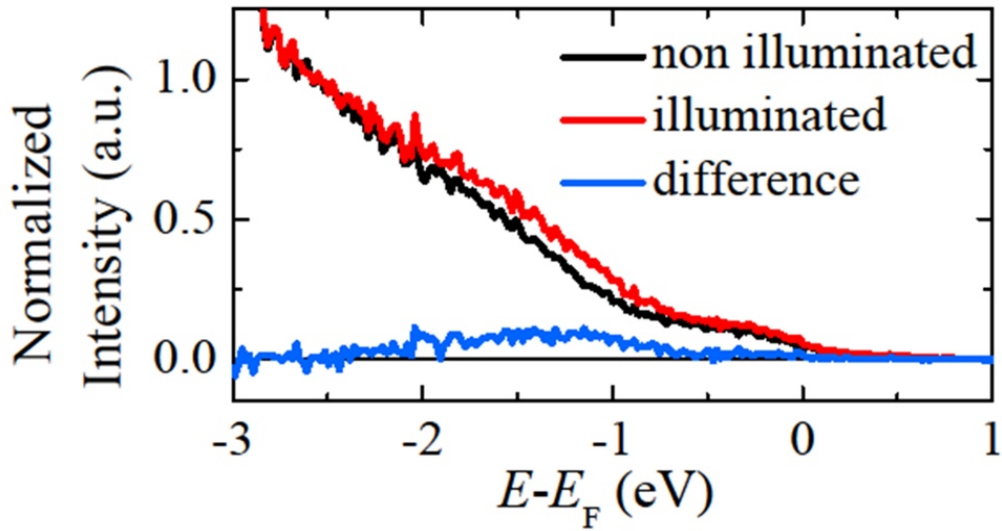

**Fig. 1 The density of states (DOS) of the $La_2NiO_{4+y}$ below the Fermi level $E_F$ before (black line, non-illuminated) and after light illumination (red line, illuminated) measured by VUV photoemission using an incident photon beam of 27 eV. A clear increase of the DOS is seen in the binding energy $E_b=E-E_F$ range centered around $E_b$=1.4 eV, illustrated by the difference curve (blue line).**

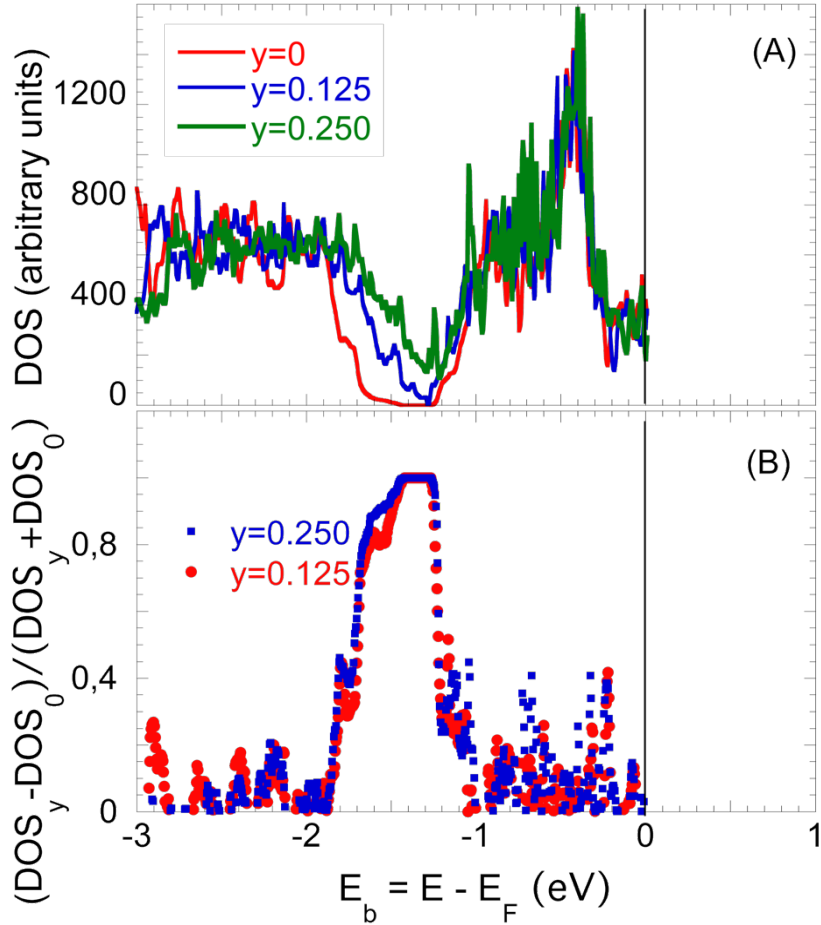

**Fig. 2 (Color online) (A) Total DOS for La$_2$NiO$_{4+y}$ for different oxygen interstitials concentrations y=0, 0.125 and 0.25 and (B) the relative increase of the DOS centered at 1.4 eV binding energy the for y=0.25 and y=0.125 oxygen interstitial concentration relative to the undoped lattice (y= 0, from reference 70).**

Energy distribution curves of selected micro-spots of illuminated and non-illuminated regions are presented in Fig. 1. A clear difference between these photoelectron spectra has been observed. In the light exposed regions an increase of the DOS compared to the non-illuminated DOS can be observed. The difference of the DOS between illuminated and non-illuminated photoelectron spectra is represented by the blue curve in Fig. 1. It reveals that states in the region around 1.4 eV below $E_F$ are those affected by light illumination. Actually, photons in the VUV region with an energy hν = 27 eV lead to a significant change of the DOS in La$_2$NiO$_{4+y}$. This result is in excellent agreement with the theoretical simulations of the increase of ordered oxygen interstitial domains presented in Fig. 2 that suggest that the DOS increases in this energy region due to oxygen interstitials.

Detailed simulations on the variation of the electronic band structure vs. oxygen interstitials concentration were performed by Jarlborg et. al. [69-71] using the linear muffin-tin orbital method and the local spin density approximation. These simulations on La$_2$CuO$_{4+y}$ are in good agreement with experimental results. The excess oxygen interstitials sit at the interstitial interlayer positions, above the oxygen ion in the CuO$_2$ plane of the orthorhombic unit cell, form 3D ordered puddles below 320 K. The non-magnetic total band DOSs for N = 0, 1 and 2 corresponding to y=0, y=0.125 and y=0.25 are displayed in Fig. 2a. It can be seen that the DOS is increasing considerably near $E_F$ when one or two Oi's are added in form of stripes, as also shown by the relative increase of the DOS in figure 2b.

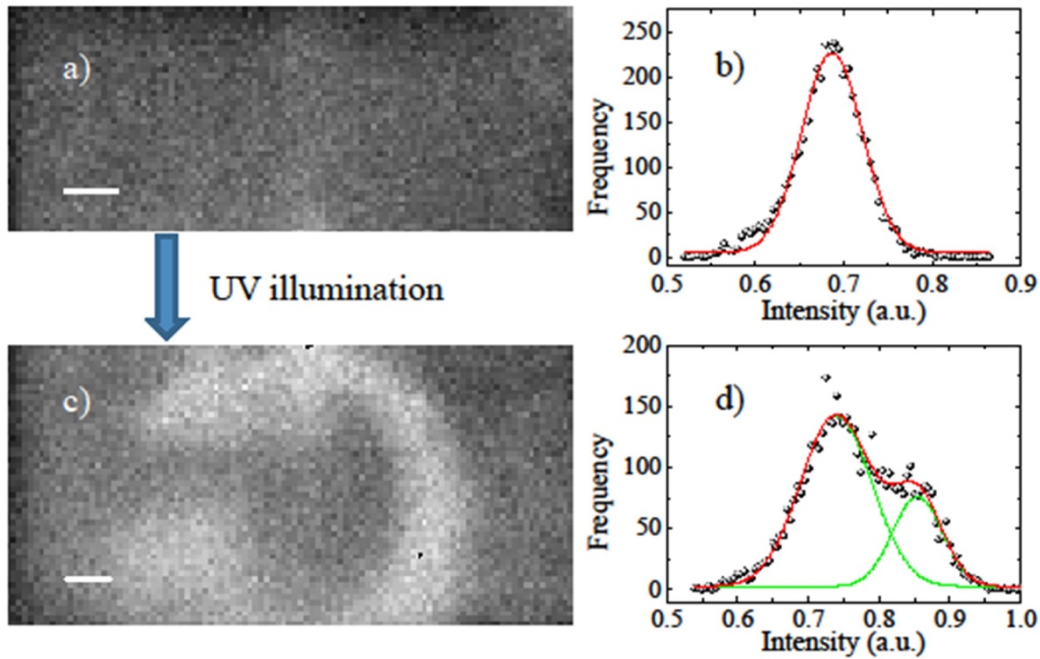

**Fig. 3 (a) Scanning photoelectron microscopy (SPEM) image of $La_2NiO_{4+y}$ of a non-illuminated region with no inhomogeneity; (b) the corresponding histogram shows a homogeneous distribution of the integrated intensity with the Gauss shape; (c) SPEM image after the illumination with the defocused beam with photons of 27 eV. Where the $La_2NiO_{4+y}$ sample was exposed to the light, an increase of the integrated intensity is observed. This is also seen in the corresponding histogram; (d) after the illumination two maxima are present representing the low-intensity non-illuminated area of the sample and the illuminated area with higher intensity. The scale bar in panel (a) and (c) corresponds to 10 μm.**

In the imaging mode, the analyzer channels well below the Fermi energy $E_F$ can be used to detect surface effects whereas others above $E_F$ are used to determine the background. This allows determining differences in the electronic structure and to spatially resolve the averaged spectroscopic information from photoemission.

In Fig. 3a is shown a SPEM image measured in the center of the Brillouin zone around Gamma. The intensity, from white to black, represents the photoemission spectroscopy yield integrated over the binding energies around 1.4 eV below $E_F$. The DOS in this binding energy range appears homogeneous over the sample surface with no significant visible features. The corresponding histogram of the normalized integrated intensity in an energy window of 0.6 eV around 1.4 eV below $E_F$ is shown in Fig. 3b. This Gauss like distribution points that the DOS at every point of the surface is the same.

The illuminated region in the SPEM image shown in Fig. 3c is represented as a bright broken circle. In this region, the DOS in the energy window of 0.6 eV around 1.4 eV below $E_F$ is higher than in the non-illuminated regions. This is demonstrated by the shape of the histogram shown in Fig. 3d that exhibits a double peak structure. The maximum of the main peak is nearly at the same integrated normalized intensity as the non-illuminated contribution shown in Fig. 3b, characteristic of the signal from the non-illuminated area of the sample. The second maximum in Fig. 3d occurs at a significantly higher integrated intensity representing the average intensity after the illumination.

## 3. CONCLUSION

In conclusion, scanning photoelectron microscopy measurements of valence band photoemission were performed in order to measure the effect of photon illumination on super-oxygenated $La_2NiO_{4+y}$ (y=0.14). The light exposure leads to an increase of the DOS mainly in the region around 1.4 eV below $E_F$, in agreement with calculations showing an increase of the DOS in the same energy window, due to oxygen interstitials. An effect of light illumination was observed by several other experiments as well. Scanning nano x-ray diffraction studies on $La_2CuO_{4+y}$ revealed that light exposure in super-oxygenated LCO lead to an ordering of the oxygen interstitials forming rows in the $La_2CuO_{4+y}$, the spacer $La_2O_2$ layer between the active layers [9-11]. Such induced ordering can be used to induce new states in transition metal oxides, supporting the development of new device possibilities. Finally this experiment shows that spectromicroscopy can be successfully used to pump and probe photoinduced mechanisms in complex solids and biological matter in quasi-stationary states out-of-equilibrium [72-76].

**Acknowledgments:** The authors thank ELETTRA, synchrotron radiation facility at Trieste, Italy for the beam time and the spectromicroscopy beamline staff for help in the experiment.

**Author Contributions:** A.B., M.B., G.C. conceived and designed the experiment; N.P. provided and characterized the single crystals; M.B. and D.I. performed the experiments; A.B., M.B., and G.C. analyzed the data; A.B., A.M., and G.C. wrote the paper.